\documentclass[a4paper,aps,prl,showpacs,amssymb,twocolumn,superscriptaddress]{revtex4}
\usepackage{graphicx}

\begin{document}
\title{Experimental evidence of stochastic resonance without tuning due to non
Gaussian noises}

\author{F. J. Castro}
\thanks{fellow of CNEA} \affiliation{Grupo Fisicoqu\'{\i}mica de Materiales}
\author{ M. N. Kuperman}
\email{ kuperman@cab.cnea.gov.ar} \affiliation{Grupo de Física
Estad\'{\i}stica
\\Centro At\'omico Bariloche and Instituto Balseiro (CNEA.and U.N.Cuyo)\\
8400--S. C. de Bariloche, Argentina}\author{ M. Fuentes}
\affiliation{Grupo de Física Estad\'{\i}stica
\\Centro At\'omico Bariloche and Instituto Balseiro (CNEA.and U.N.Cuyo)\\
8400--S. C. de Bariloche, Argentina}
\author{H. S. Wio}
\email{wio@cab.cnea.gov.ar.} \affiliation{Grupo de Física
Estad\'{\i}stica
\\Centro At\'omico Bariloche and Instituto Balseiro (CNEA.and U.N.Cuyo)\\
8400--S. C. de Bariloche, Argentina}

\begin{abstract}
In order to test theoretical predictions, we have studied the phenomenon of
stochastic resonance in an electronic experimental system driven by white
{\bf non Gaussian} noise. In agreement with the theoretical predictions our
main findings are: an enhancement of the sensibility of the system together
with a remarkable widening of the response (robustness). This implies that
even a single resonant unit can reach a marked reduction in the need of
noise tuning.
\end{abstract}


\maketitle
\newpage

The phenomenon of {\it Stochastic resonance} (SR) has attracted enormous
interest due to both its potential technological applications for optimizing
the response to weak external signals in nonlinear dynamical systems and its
connection with some biological mechanisms. A recent review shows the state
of the art \cite{srrmp}. There, a large number of applications are shown in
science and technology, ranging from paleoclimatology \cite{BEN}, to
electronic circuits \cite{elect1,elect2}, lasers \cite{laser}, chemical
systems \cite{chem}, and the connection with some situations of biological
interest (noise-induced information flow in sensory neurons in living
systems, influence in ion-channel gating or in visual perception) \cite{biol}%
. Recent works have shown a tendency pointing towards achieving an
enhancement of the system response (by means of the coupling of several SR
units in what conforms an {\it extended medium} \cite{buls1,nosot,bowi}), or
analyzing the possibility of making the system response less dependent on a
fine tuning of the noise intensity \cite{claudio}, as well as different ways
to control the phenomenon \cite{control}.

It is worth remarking here that a majority of such studies on SR, with very
few exceptions \cite{nG}, have been done assuming that the noises were
Gaussian. However, some experimental results in sensory systems,
particularly for one kind of crayfish \cite{circle} as well as recent
results for rat skin \cite{nuevo}, offer strong indications that the noise
source in these systems could be non Gaussian. Some recent studies in neural
networks also point in this direction \cite{German}.

In recent works \cite{NG1,NG2}, a nice way of generating a non Gaussian
noise with a ``fine'' control on the degree of non Gaussianity was
introduced and some interesting results concerning the effect of using non
Gaussian noises in the study of SR have been obtained. The results of some
analytical approximations and of numerical simulations indicate a certain
degree of enhancement of the system response when it departs from Gaussian
behavior. However, a most remarkable finding was that the system shows a
marked ``robustness'' against noise tuning. Such robustness means that the
maximum of the signal-to noise ratio (SNR) curve can flatten when departing
from Gaussian behavior, implying that the system does not require a fine
tuning of the noise intensity in order to maximize its response to a weak
external signal.

In this letter we analyze the case of SR when the noise source is non
Gaussian but from an experimental point of view. We have studied an
experimental set up similar to the one used in \cite{elect1}, but now using
a non Gaussian noise source which was built exploiting the form of noise
introduced in \cite{NG1,NG2}, but for the particular case of white noise. In
\cite{NG1,NG2} a particular class of Langevin equations was studied, having
non Gaussian stationary distribution functions \cite{borland1}. The work in
\cite{borland1} is based on the generalized thermostatistics proposed by
Tsallis \cite{TS} which has been successfully applied to a wide variety of
physical systems \cite{TS2}. Those equations were
\begin{eqnarray}
\dot{x} &=&f(x,t)+g(x)\eta (t)  \label{equis} \\
\dot{\eta} &=&-\frac{1}{\tau }\frac{d}{d{\eta }}V_{q}(\eta )+\frac{1}{\tau }
\xi (t)  \label{nu}
\end{eqnarray}
where $\xi (t)$ is a Gaussian white noise of zero mean and correlation $<\xi
(t)\xi (t^{\prime })>=2D\delta (t-t^{\prime })$, and $V_{q}(\eta )$ is given
by \cite{borland1} $V_{q}(\eta )=\frac{1}{\beta (q-1)}\ln [1+\beta (q-1)
\frac{\eta ^{2}}{2}]$, where $\beta =\frac{\tau }{D}$. The function $f(x,t)$
was derived from a potential $U(x,t)$, consisting of a double well potential
and a linear term modulated by $S(t)\sim F\,\cos (\omega t)$ ($f(x,t)=-\frac{%
\partial U}{\partial x}=-U_{0}^{\prime }+S(t)$). This problem corresponds
(for $\omega =0$) to the case of diffusion in a potential
$U_{0}(x)$, induced by $\eta $, a colored non-Gaussian noise.
Clearly, when $q\to 1$ we recover the limit of $\eta $ being a
Gaussian colored noise (Ornstein-Uhlenbeck process). The
stationary probability distribution for the random variable $\eta
$ and $q>1$ is given by $P_{q}^{st}(\eta )=\frac{1 }{Z_{q}}\left[
1+\beta (q-1)\eta ^{2}/2\right] ^{\frac{-1}{q-1}}$, with $\eta \in
(-\infty ,\infty )$, and $Z_{q}$ the normalization factor. When
$q<1$ the expression adopts the form:
\begin{equation}
\begin{array}{ll}
P_{q}(\eta )= & \left\{
\begin{array}{cc}
\frac{1}{Z_{q}}\left[ 1+\beta (q-1)\eta ^{2}/2\right]
^{\frac{-1}{q-1}} &
\text{ if }|\eta |<w \\
&  \\
0 & \text{otherwise,}
\end{array}
\right.
\end{array}
\nonumber
\end{equation}
with $w^{-2}=(1-q)\beta ,$ and with noise intensity $D=2 \beta
^{-1}/(5-3q)$.

By applying a path-integral formalism to the Langevin equations given by
Eqs. (\ref{equis}),(\ref{nu}), and making an adiabatic-like elimination
procedure \cite{pi1} it was possible to arrive at an {\it effective
Markovian approximation}. The specific details, not relevant here, have been
shown elsewhere \cite{NG1}. In the present work we have used a stationary
probability distribution for the noise with the same form indicated above,
and for $q<1.$ To obtain a number with the mentioned probability
distribution we generate two uniformly distributed random numbers $x$ $\in
(-w,w)$ and $y\in (0,P_{q}(0)).$ $x$ is accepted if $y<P_{q}(x).$

To make the experiment we have used an electronic (analogical) circuit. We
have used the well known Schmitt trigger, in an arrangement similar to \cite
{elect1}. Analogical approaches have been extensively used in order not only
to study SR (see Section II.C of \cite{srrmp}) but also to analyze the
propagation of freezing fronts, spatiotemporal SR, irreversibility of
classical fluctuations, SR in nonlinear electronic systems, resonant
propagation, etc. \cite{electvar}.

The experimental setup was implemented with a LM741 integrated opamp with
positive feedback and variable resistors in a voltage divider configuration
that could be used to set appropriately the threshold level of the circuit.
The input and output signals were driven by a 12 bit data acquisition board
(Advantech PCL-818L) connected to a PC. The frequencies of the periodic
signal provided to the circuit was chosen as 100 Hz or 250 Hz, with an
amplitude that could be varied between 0 and 3.536 $V_{\mbox{rms}}$. The
threshold level of the circuit was set as 3.018 $V_{\mbox{rms}}.$ The
optimal temporal resolution achieved with the system was 70 $\mu $s. The
noises were obtained using a numerical generator as indicated above.

The set of parameters controlled during the experiments were the amplitude ($%
V_{s}$) and frequency ($\nu $) of the periodic signal, the noise intensity $%
D $ and the parameter $q$ of the noise distribution. For each election of
the parameter values we performed measurements comprising 300000 points with
a time lapse of 70 $\mu $s between two consecutive acquisitions.. The
phenomena of SR was characterized by the residence time distribution \cite
{srrmp}. In particular, we considered the area ($A_{\nu }$) and the height ($H_{\nu }$) of the first peak of the distribution, centered at $t=1/(2\nu )$.
In Figs. 1 and 2 we display curves showing the behavior of $A_{\nu }$ as a
function of $D$, for several situations.

Figure 1 shows the effect of varying the $q$ parameter for $\nu =100$ Hz and
$V_{s}=1.414V_{\mbox{rms}}$. We can see that for each value of $q$, there is
a clear resonant noise intensity. As $q$ decreases we find that the resonant
noise intensity shifts to higher $D\;$values and the region where $A_{\nu }$
remains high is much wider (robustness). At the same time, the value of the
maximum of each curve is higher. These results are in correspondence with
previously reported ones obtained from numerical simulations \cite{NG2}. In
Fig. 2 we show the effect of changing the amplitude $V_{s}$ and the
frequency $\nu $ of the periodic signal. We can see that the resonant range
for $D$ turns wider and the sensitivity of the system increases for growing $V_{s}$. On the other hand, the effect of increasing the frequency is to
shift the resonant peak towards higher values of $D$.

As mentioned earlier, $A_{\nu }$ corresponds to the area under the first peak of the residence time
distribution. This quantity produces neat curves that allow us to show more
clearly the SR effect. However, if we want to compare the effect of $q$ on
the sensitivity of the system it is more convenient to plot the height of
the peak, $H_{\nu }$, for a chosen value of $V_{s}$ and $\nu ,$ and several $q$ values. The latter is shown in Fig. 3 where we can see that there is a
clear enhancement of the sensitivity of the system with decreasing $q$.

Summarizing, motivated by some experimental results in sensory systems \cite
{circle,nuevo}, as well as some recent theoretical results \cite{NG1,NG2},
we have experimentally analyzed the problem of SR when the noise source is
non Gaussian. We have chosen a white non Gaussian noise source with a
probability distribution based on the generalized thermostatistics \cite{TS}.

Our experimental results indicate that

\begin{itemize}
\item  The resonance range increases significantly with the non-gaussianity
of the noise. That shows that tuning the noise intensity is not determinant
in order to increment the signal perception. This effect is what we call
increased robustness of the system.

\item  At the same time there is a clear enhancement of the sensitivity of
the system (rising of $H_{\nu }$) with decreasing $q.$

\item  For larger amplitude ($V_{s}$) of the periodic signal, the
``resonant'' range for $D$ becomes wider and the height of the response
increases.

\item  The effect of increasing the frequency $\nu $ is to shift the
resonance peak towards lower values of $D$.
\end{itemize}

As we depart from Gaussian behavior (with $q<1$), the SNR shows two main
aspects: firstly it becomes less dependent on the precise value of the noise
intensity, and secondly its maximum as a function of the noise intensity
increases. Both aspects are of great relevance for technological
applications \cite{srrmp}. Moreover, as was indicated in \cite{nuevo}, non
Gaussian noises could be an intrinsic characteristic in biological systems,
particularly in sensory systems \cite{biol,circle,nuevo}. In addition to the
increase in the response (SNR), the reduction in the need of {\it tuning} a
precise value of the noise intensity is of particular relevance both in
technology and in order to understand how a biological system can exploit
this phenomenon. It is worth remarking here that such effects have been
obtained considering a single resonant unit and that, according to the
results of previous works \cite{buls1,nosot,bowi}, one should expect an even
larger effect when several units are coupled. Here we have focused on the
case of white non Gaussian noise, the case of colored non Gaussian noise
will be the subject of a forthcoming work.\newline

\noindent {\bf Acknowledgments:} The authors thank J. Bergaglio for
technical assistance and V. Grunfeld for a revision of the manuscript, and
acknowledge support from CONICET, Argentine. M.K. and M.F. thank
Fundaci\'{o}n Antorchas for partial support.


\noindent {\bf FIGURE CAPTIONS}\newline

\noindent Figure 1: $A_{\nu }$ vs. $D$ for several $q $ values. $\nu =100$Hz
and $V_{s}=1.414$ $V_{\mbox {rms}}$.(solid)$\,q=1$; (dashed) $\,q=0.5$;
(dotted) $\,q=0.25$; (dash-dotted) $\,q=0.1$. \newline

\noindent Figure 2: $A_{\nu }$ vs. $D$ for $q=0.1$ and varying $V_{s}$ and $%
\nu :$ (dotted)$\nu =100$ Hz and $V_{s}=2.121$ $V_{\mbox {rms}}$; (dashed) $%
\nu =250$ Hz and $V_{s}=1.414$ $V_{\mbox
{rms}}$; (solid) $\,\nu =100$ Hz and $V_{s}=1.414$ $V_{\mbox
{rms}}$.\newline

\noindent Figure 3: $H_{\nu }$ vs. $q$, for $\nu =100 $Hz and $V_{s}=1.414$ $V_{\mbox {rms}}$.\newline

\end{document}